\documentclass[12pt]{article}
\usepackage{amssymb,bbold}
\usepackage{graphicx}
\usepackage[hang,small,bf]{caption}
\newcommand{\be}[3]{\begin{equation}  \label{#1#2#3}}
\newcommand{\ee}{\end{equation}}
\newcommand{\ba}{\begin{array}}
\newcommand{\ea}{\end{array}}
\newcommand{\bea}[3]{\begin{eqnarray}  \label{#1#2#3}}
\newcommand{\eea}{\end{eqnarray}}

\let\Large=\large
\let\large=\normalsize

\def\1{\mathbb 1}

\newcommand{\haken}{\mathbin{\hbox to 8pt{%
                 \vrule height0.4pt width7pt depth0pt
                 \kern-.4pt
                 \vrule height4pt width0.4pt depth0pt\hss}}}

\begin{document}

\baselineskip=19pt
\parskip=6pt


\thispagestyle{empty}

\begin{flushright}
\hfill{AEI-2003-111}\\
\hfill{HU-EP-04/01} \\
\hfill{hep-th/0401019}

\end{flushright}

\vspace{10pt}

\begin{center}{ \Large{\bf
Superpotentials from flux compactifications of
M-theory\footnote{Contribution to proceedings of the "36th
International Symposium Ahrenshoop on the Theory of Elementary
Particles," Berlin, August 26-30, 2003; and the RTN-workshop ``The
quantum structure of space-time and the geometrical nature of the
fundamental interactions'', Copenhagen, September 15-20, 2003.}  }}

\vspace{35pt}

{\bf Klaus Behrndt}$^a$ \quad and \quad
{\bf Claus Jeschek}$^b$

\vspace{15pt}

$^a$ {\it  Max-Planck-Institut f\"ur Gravitationsphysik,
Albert Einstein Institut\\
Am M\"uhlenberg 1,  14476 Golm,
 Germany}\\[1mm]
{\tt E-mail: behrndt@aei.mpg.de}

\vspace{8pt}

$^b$ {\it  Humboldt Universit\"at zu Berlin,
Institut f\"ur Physik,\\
Newtonstrasse 15, 12489 Berlin, Germany }\\[1mm]
{\tt E-mail: jeschek@physik.hu-berlin.de}

\vspace{40pt}

{ABSTRACT}

\end{center}

\noindent
In flux compactifications of M-theory a superpotential is generated
whose explicit form depends on the structure group of the
7-dimensional internal manifold.  In this note, we discuss
superpotentials for the structure groups: G$_2$, SU(3) or SU(2). For
the G$_2$ case all internal fluxes have to vanish.  For SU(3)
structures, the non-zero flux components entering the superpotential
describe an effective 1-dimensional model and a Chern-Simons model if
there are SU(2) structures.

\vfill

\newpage

In order to make phenomenological predictions one has to fix the
moduli appearing in string or M-theory compactifications. One way of
doing this is to consider flux compactifications, where the generated
superpotential might fix most of the moduli, although it is not clear
whether all moduli can be fixed.  The resulting potentials have not
only supersymmetric extrema, related to anti deSitter (AdS) or
preferable flat space vacua, but may also have de Sitter vacua which
are interesting in its own.

{From} the Killing spinor equations we get not only constraints on the
fluxes and the geometry of the internal space, see \cite{111, 110,
130}, but also, as we will present in this note, a procedure to
calculate the superpotential.  In the usual way one derives the
superpotential by dimensional reduction or from calibrated
submanifolds \cite{131,200,132}, but we will follow another route. We
introduce the superpotential as mass term for the 4-d gravitino(s),
implying that the four-dimensional (4-d) Killing spinor is not
covariantly constant, and hence the 11-d Killing spinor equations
relate directly the lower-dimensional superpotential to the fluxes.

By definition, in the vacuum all 4-d scalars as well as gauge fields
are trivial and hence the metric is either flat or AdS. We write
therefore the 11-d bosonic fields as
\be012
\ba{rcl}
ds^2 &=& e^{2 A} \Big[ \, g^{(4)}_{\mu\nu} dx^\mu dx^\nu
        + \, h_{ab} dy^a dy^b \Big] \ , \\[4mm]
F &=& m \,  
dx^0 \wedge dx^1 \wedge
    dx^2 \wedge dx^3 +
{1 \over 4!} F_{abcd} \, dy^a \wedge dy^b \wedge dy^c \wedge dy^d
\ea
\ee
where $A=A(y)$ is a function of the coordinates of the internal
7-manifold $X_7$ with the metric $h_{ab}$ and we denote the Freud-Rubin
parameter as $m$ and $g^{(4)}_{\mu\nu}$ is the 4-d metric.
Unfortunately, we have to omit here many technical details which are
important for the understanding and refer only to the literature
\cite{130}.

Unbroken supersymmetry requires the existence of (at least) one
Killing spinor $\eta$ yielding a vanishing gravitino variation of
11-dimensional supergravity
\be716
\ba{rcl}
0 = \delta \Psi_M &=& 
        \Big[ \partial_M + {1 \over 4} \hat \omega^{RS}_M \Gamma_{RS}
        + {1 \over 144} \Big(\Gamma_M^{\ NPQR} - 8 \, \delta_M^N\, 
        \Gamma^{PQR} \Big) \, F_{NPQR} \Big] \eta \\[4mm]
&=& 
        \Big[ \partial_M + {1 \over 4} \hat \omega^{RS}_M \Gamma_{RS}
        + {1 \over 144} \Big(\Gamma_M \hat F - 12 \, \hat F_M \Big) \Big] \eta 
\ .
\ea
\ee
In the second line we used the formula: $\Gamma_M \Gamma^{N_1 \cdots
N_n} = \Gamma_M^{\ N_1 \cdots N_n} + n \, \delta_M^{\ [N_1}
\Gamma^{N_2 \cdots N_n]}$ and introduced the abbreviation $\hat F
\equiv F_{MNPQ} \Gamma^{MNPQ}$ , $\hat F_M \equiv F_{MNPQ}
\Gamma^{NPQ}$. We decompose the $\Gamma$-matrices
as usual
$
\Gamma^\mu = \hat \gamma^\mu \otimes 1$ , $
\Gamma^{a+3} = \hat \gamma^5 \otimes \gamma^a
$
with $\mu = 0,1,2,3$, $a = 1,2, \ldots 7$ and we find for the field strength
\be018
\hat F = - i \, m \, \hat \gamma^5 \otimes {\bf 1} + 
{\bf 1} \otimes F\ , \qquad 
\hat F_\mu = {1 \over 4} \, i \, m  \hat \gamma^5 \hat \gamma_\mu \otimes 
   {\bf  1} \  , \qquad  \hat F_a = \hat \gamma^5 \otimes F_a \ .
\ee
One distinguishes the external and internal variation, where the internal
variation gives a differential equation fixing not only the
spinor but giving also differential equations for the metric (or the
vielbeine). Its variation reads
\be836
0 = \Big[ {\bf 1} \otimes \Big( \nabla_a^{(h)} + {1 \over 2}
  \gamma_a^{\ b} \, \partial_b A + {i\, m  \over 144} \, \gamma_a \Big) + 
        {1 \over 144} e^{-3A}\,  \hat \gamma^5 \otimes
  \Big( \gamma_a F  -12 F_a \Big) \Big] \eta \ .
\ee
In this paper we will explore especially the external variation which
becomes
\be112
0=\Big[ \nabla_\mu \otimes 1 +  \hat \gamma_\mu \hat \gamma^5
  \otimes \Big( {1 \over 2} \, \partial A + {i \, m \over 36} \Big)
+ {1 \over 144} e^{-3A} \, \hat \gamma_\mu \otimes F
  \Big] \eta  
\ee
where $\partial A \equiv \gamma^a \partial_a A$ and $\nabla_\mu$ is
the 4-d covariant derivative in the metric $g^{(4)}_{\mu\nu}$.

As next step one has to expand the 11-d Majorana spinor in all
independent 4-d spinors $\epsilon^i$ and 7-d (real) spinors $\theta_i$
giving
\be628
\eta = \sum_{i=1}^N \, \epsilon^i \otimes \theta_i  \equiv
\epsilon^i \otimes \theta_i  \ .
\ee
If there are no fluxes, all of these spinors are covariantly constant
and $N$ gives the number of extended supersymmetries in 4 dimensions,
with $N=8$ as the maximal supersymmetric case. If one turns on fluxes,
not all spinors are independent resulting in a reduction of the number
of preserved supersymmetries.  In the least supersymmetric case, we
have only a single spinor on $X_7$, which at each point of the
internal space can be written as a singlet of $G_2 \subset SO(7)$, but
in general the embedding changes from point to point. In fact the
corresponding rotation is related to the structure group $G \subset
G_2$ and the spinor is only a singlet under the structure group, but
in general not under G$_2$. One expands now the 11-d spinor $\eta$
with respect to the maximal number of singlets under the structure
group, i.e.\ for G$_2$: $N$=1 ; SU(3): $N$=2 ; $Sp(2)$: $N$=3 and for
SU(2): $N$=4. Recall, only for very specific fluxes this agrees with
the number of supersymmetries and for generic fluxes we will always
get constraints on the spinors so that we encounter in general an
$N$=1 vacuum.

Before we come to a discussion of the different cases, we have to
introduce the superpotential, which by definition is the mass term of
the 4-d gravitino(s). This implies that the 4-d spinor(s) are not
covariantly constant but satisfy the equation (for simplicity we put
all numeric factors into the superpotential)
\be811
\nabla_\mu \epsilon^i \  = \ \hat \gamma_\mu \, 
 ( W_1^{ij} + i \hat \gamma^5 \,W_2^{ij} ) 
\, \epsilon_j \ .
\ee
If $N$ is even one may introduce a symplectic Majorana notation, but
in order to cover also the case with a single spinor let us stick for
the time being to the real notation.  This yields for the external
variation
\be625
0  = \Big( [ W_1^{ij} + i \hat \gamma^5 \,W_2^{ij} ]\otimes {\bf 1}  
+ \delta^{ij} \Big[\, {i m \over 36} \,\hat \gamma^5 \otimes {\bf 1}  +
 {1 \over 2} \, \hat \gamma^5 \otimes \partial A + {1 \over 144}
\,e^{-3A}\, ({\bf 1} \otimes F) \Big] \, \Big) \epsilon_i \otimes \theta_j  \ .
\ee
Note, $W^{ij}_{1/2}$ mixes the different components of $\epsilon^i$
and in general one has to impose constraints on the 4-d spinors to
solve these equations.  Let us now consider the different cases
separately.

\bigskip

\noindent
{\em \large Case (i): G$_2$-structure} 

\medskip

\noindent
If the structure group is the whole G$_2$, only one real spinor on
$X_7$ can be singlet. Hence the 11-d spinor is written as
\be251
\eta = \epsilon \otimes \theta
\ee
and since the 11- and 7-d spinor are Majorana also the 4-d spinor
$\epsilon$ has to be Majorana.  In this case the spinor $\theta = e^Z
\theta_0$ (with the real function $Z$) is a G$_2$ singlet, i.e.\ the
constant spinor $\theta_0$ obeys:
$
\gamma_{ab} \theta_0 = - i \varphi_{abc} \gamma^c \theta_0\ ,
$
with the G$_2$ invariant 3-index tensor $\varphi_{abc}$ which in turn
can be defined as fermionic bi-linear $\varphi_{abc} = -i \theta_0^T
\gamma_{abc} \theta_0$ (note: $\theta_0^T \gamma_a \theta_0 =
\theta_0^T \gamma_{ab}\theta_0 = 0$). In our notation, the 7-d
$\gamma$-matrices are purely imaginary and the $\theta_0 =
(1,0,0...)$.

Now, inserting this spinor ansatz into (\ref{625}) we get two
equations; one proportional to $\epsilon$ and another proportional to
$\hat \gamma^5 \epsilon$. Contracting them $\theta$ gives
\bea440
W_1 &=& {1 \over 144} e^{-3A} (\theta^T F \theta) = {1 \over 144} \,
 e^{-3A+2Z} \, \psi^{abcd} F_{abcd}\ ,  \\  \label{874}
W_2 &=& {m \over 36} 
\eea
where the 4-index tensor $\psi_{abcd}$ as the dual of $\varphi_{abc}$.
This superpotential can be written in the form proposed in \cite{132}:
$W \sim \int F \wedge (\varphi + {i \over 2} C)$ if one uses the
equations of motion for the gauge field: $ d(\ast_{11} F + { 1\over 2}
F \wedge C) =0$ for the external components, where $\int_{X_7}
\ast_{11} F = \int_{X_7} F_7 \sim \int_{X_7} \ast_7 m$.  

But, since we are dealing with a Majorana spinor, also the internal
variation (\ref{836}) gives two independent equations: one
differential equation ($\sim \epsilon$) and another constraints on the
fluxes ($\sim \hat \gamma^5 \epsilon$).  The latter implies that
$\psi^{abcd} F_{abcd} =0$ and hence only the Freud-Rubin parameter $m
\sim W_2$ can be non-zero. In fact a detailed analysis shows that all
internal fluxes have to be trivial in this case \cite{233, 130}.

\bigskip

\noindent
{\em \large Case (ii): SU(3)-structure} 

\medskip

\noindent
Next, if we reduce the structure group to SU(3), one can build two
singlet spinors on $X_7$. These two spinors are equivalent to the
existence of a vector field $v$, which in turn can be expressed as a
bi-linear of these two spinors. If we normalize this vector field,
these two real spinors can be combined into one complex spinor defined
by
\be726
\theta = {1 \over \sqrt{2}} \, e^Z \, ( 1 + v_a \gamma^a ) \theta_0
\quad , \qquad v_a v^a = 1
\ee
where the constant spinor $\theta_0$ is again the G$_2$ singlet and
$Z$ is now a complex function. The vector field defines a foliation of
the 7-manifold by a 6-manifold $X_6$ and both spinors, $\theta$ and
its complex conjugate $\theta^\star$, are chiral spinors on $X_6$. 
The 11-d Majorana spinor is now decomposed as
\be010
\eta = \epsilon \otimes \theta + \epsilon^\star \otimes \theta^\star 
\ee
where the complex 4-d spinor is chiral and we choose
\[
\hat \gamma^5 \epsilon = \epsilon \quad , \qquad \hat \gamma^5 \epsilon^\star
= - \epsilon^\star \ .
\]
We introduce the superpotential by
\[
\nabla_\mu \epsilon \ = \
\hat \gamma_\mu \, e^{K \over 2} \, \bar W \, \epsilon^\star
\]
with $\bar W = W_1 - i \, W_2$ and $K$ is the K\"ahler potential [we
use here the known notation from 4-d supergravity]. Due to the
opposite chirality, terms with $\epsilon$ and $\epsilon^\star$ are
independent and the term ${\cal O}(\epsilon)$ of the external
variation (\ref{625}) can be written as
\be626
e^{K \over 2}\, W  \, \theta^\star  = -  \Big[\, {i m \over 36} \,  +
 {1 \over 2} \, \partial A + {1 \over 144}
\,e^{-3A}\, F \Big] \, \theta  \ .
\ee
Again, we can contract this equation with $\theta^T$ and use $\theta^T
\theta = \theta^T \gamma_a \theta = 0$ ($\theta^+ \theta = e^{Z + \bar
Z}$) to find 
\be234
e^{K \over 2}\, \bar W = {1 \over 4! \, 3!} e^{-(3A + Z+ \bar Z)} \, 
(\theta^T F \theta) =  {i \over 36 }\, 
e^{-3A} \, v^a F_{abcd} \bar \Omega^{bcd}
\ee
where $\Omega$ is a holomorphic 3-form on $X_6$. If we define on $X_6$
the 3-form field strength by $ H_{abc} \equiv v^d F_{dabc}$ and if we
integrate over the internal space, the superpotential can also be
written as (note the volume form on $X_6$ is: $i \, \Omega \wedge \bar
\Omega$)
\be832
W \ \sim \ \int_{X_7}  H \wedge \Omega \wedge v 
\ \sim \ \int_{X_7} F \wedge \Omega \ .
\ee
Let $\omega$ be the associated 2-form to the almost complex structure
on $X_6$ defined by: $\omega_{ab} = \varphi_{abc} v^c$.  One finds
that: $d \omega \wedge \Omega \sim W \, i\, \Omega\wedge \bar \Omega$
\cite{130}, which allows to express the superpotential also purely
geometrically in terms of torsion components. {From} our setup we
cannot distinguish between both expressions, but if $v$ is Killing, we
can compare our result with the ones derived in \cite{244} and this
suggest that we have to add both expressions yielding
\be372
W  \ \sim \  \int_{X_7} (F + {i} v \wedge d \omega)\wedge \
\Omega^{(3,0)} \ .
\ee
Upon reduction to 10 dimensions, this superpotential contains only
fields that are common in all string theories and hence one can make a
number of consistency checks (that it is U-dual to the type IIB
superpotential and anomaly-free on the heterotic side).

Let us continue and contract eq.\ (\ref{626}) with $\theta^+$ and find
\[
0= - { i m \over 36} + {1 \over 2} v^a \partial_a A + 
{1 \over 144} e^{-3A}\, (\theta^+ F \theta) \ .
\]
Since the last term is real, we are forced to set $m= 0$ and $v^a
\partial_a A = {e^{-3A} \over 144} (F_{abcd} \omega^{ab}
\omega^{cd})$. If one reduces as first step only over $X_6$, the
corresponding 5-d superpotential becomes $W_{5d} = (F_{abcd}
\omega^{ab} \omega^{cd}) \sim \int_{X_6} F \wedge \omega$.  Note, this
$W_{5d}$ is compensated by the warp factor $A$ and does not enter the
4-d superpotential. In fact, from the supergravity point of view one
obtains in 5 dimensions a domain wall solution and this first order
differential equation for $A$ is the known BPS equation, see
\cite{200}.  The superpotential $W$ on the other hand does not
contribute to the 5-d potential, but represents a kinetic term of two
(axionic) scalars of a hyper multiplet\footnote{These scalars come
from the (3,0) and (0,3) part of the 3-form potential $C$ in 11
dimensions.}, which is non-zero in the vacuum and curves the domain
wall.  That kinetic terms of (axionic) scalars act effectively as
potentials (if one allows for a linear dependence of the $5^{th}$
direction) can be understood from massive T-duality or generalized
dimensional reduction \cite{332} and hence we expect that these flux
compactifications are related to the curved domain wall appeared in
\cite{430}. Let us also note that, because we excluded any dependence
on the four external coordinates, this reduction over $X_6$ gives an
effective 1-dimensional description and if one would take into account
singularities giving rise to non-Abelian gauge groups, we expect a
matrix model description as discussed in \cite{028}.

\bigskip

\noindent
{\em \large Case (iii): SU(2)-structure} 

\medskip

\noindent
Strictly speaking the reduction of the structure group would yield as
next step the group $SO(5)\simeq Sp(2)$, related to three (real)
singlet spinors on $X_7$. We will however, not discuss this case here,
which for trivial fluxes is related to $N=3$ models. Instead, we will
consider the case where the structure group is SU(2), which allows for
four singlet spinors and this case appears as natural continuation of
the SU(3) case. It is again natural to work with (two) 4-d chiral
spinors and we write the 11-d spinor as
\be011
\eta = \epsilon^i \otimes \theta_i + cc \quad , 
\qquad {\rm with}: \quad \hat \gamma^5 \epsilon^i = \epsilon^i
\ee
where $i = 1,2$. Similarly we combine the four real spinors on $X_7$
into two complex spinor defined by
\[
\theta_1 = {1 \over \sqrt{2}} \, e^Z ( 1 + v_a \gamma^a) \theta_0 \ , \quad
\theta_2 = {1 \over \sqrt{2}} \, e^Y ( u_a + i \,w_a) \gamma^a\, \theta_0 
= e^{Y -Z}\,  u_a \gamma^a \, \theta_1
\]
with $Z$ and $Y$ as complex functions and the three vectors are
orthogonal and normalized: $|v|^2 = |u|^2 =|w|^2$, $u \cdot w = u
\cdot v= w \cdot v =0$ and obey moreover the relation $w_a =
\varphi_{abc} v^b u^c$.  In addition, they can be expressed as
fermionic bi-linears
\[
v_a = e^{-(Z + \bar Z)} \, (\theta_1^+ \gamma_a \theta_1) \quad , \qquad 
u_a + i w_a \, = \, e^{-(\bar Z + Y)}\, (\theta_1^+ \gamma_a \theta_2) \ .
\]
and to simplify the notation we will in the following set $Y=Z=0$.
These three vectors imply that the 7-manifold $X_7$ is a fibration of
a 3- over a 4-manifold $X_4$, which is reminiscent to the known
G$_2$-manifolds written as $R_3$-fibrations over a self-dual Einstein
space or an $S^3$ fibration over a hyper-K\"ahler space
\cite{032,006}.  The concrete geometry is fixed by the differential
equations satisfied by these vector fields, which however is not the
subject of this paper.  Let us continue and note, that the vector $v$
can be used to define an SU(3) structure and the holomorphic
vector\footnote{We use holomorphicity in a pointwise sense, i.e.\
using $\varphi_{abc}v^c$ as an almost complex structure on $X_6$, we
can at each point introduce holomorphic indices and the vector, e.g.,
obeys: $(1-i \, \varphi_{abc}v^c) (u^a + iw^a)=0$, because $w_a =
\varphi_{abc}v^b u^c$. Note, in general $X_6$ is {\em not} a complex
manifold.}  $u+iw$ gives the reduction to SU(2).  In this case, the
external variation can be written as
\be685
e^{K \over 2} \,  W^{ij} \, \theta^\star_j =  - \Big[\, {i m \over 36}   +
 {1 \over 2} \, \partial A + {1 \over 144}
\,e^{-3A}\, F \Big] \,  \theta^i   \ .
\ee
As for the SU(3)-case, we had to collect again terms of the same
chirality and the superpotential has been promoted from a complex to
an SU(2)-valued quantity, which together with the K\"ahler potential
transforms under the U(2) R-symmetry rotation of the two spinors.
It can be written as
\be629
W^{ij} \equiv \, W_x (\sigma^x)_k^{\ j} \epsilon^{ik}
= -i \, W_x (\sigma^x \cdot \sigma_2)^{ij} 
\ee
where $\sigma^x$ are the three Pauli matrices (this notation is again
borrowed from N=2 gauged supergravity, see \cite{012,190}) and the
three components $W_x$ can be obtained by contracting eq.\ (\ref{685})
by $\theta_i^T$. Because $\theta_i^+ \theta_j = \theta_i^T
\theta^\star_j =\delta_{ij}$ and in addition $\theta^T_i \theta_j =
\theta^T_i \gamma_a \theta_j =0$ we find
\be829
W_{ij} = {1 \over 144} e^{- {K \over 2} -3A}\, ( \theta^T_i F \theta_j) \ .
\ee
Since $\epsilon^{ij} \theta_i^T F \theta_j \sim \theta_1 \{F, u\} \theta_1
=0$, this matrix is symmetric which is in agreement with
(\ref{629}). Using the Pauli matrices we can write
\be847
W  \sim \int_{X_7} F \wedge q \wedge \Omega^{(2)}
\qquad {\rm with:} \quad q = u \, {\bf \sigma_3} -i w \, {\bf 1} 
 - v \, \sigma_1 
\ee
where $\Omega^{(2)}_{ab} = \Omega_{abc}(u^c + i w^c) =2 \,
\Omega_{abc}u^c$ is the holomorphic 2-form on $X_4$.  In order to
ensure that $W_x$ as introduced in (\ref{629}) are real we impose that
the three 2-forms $\hat F_2 = [{}^\star (F \wedge v), {}^\star (F
\wedge u),{}^\star (F \wedge w)]$ have no components along ${\rm Im}
\Omega^{(2)}$ on $X_4$.  Note there are three 2-forms on $X_4$ which
are SU(2) singlets and which one may choose as anti self-dual: two of
them are combined in the holomorphic $\Omega^{(2)}$ and the third one
is the complex structure. For generic fluxes they might not be exact
and similar to the situation with SU(3) structures, we expect here
additional contributions to the superpotential, which have not yet
been worked out. 

Also in this case, the superpotential can
effectively be described by a lower dimensional model. We can again
ignore the four external coordinates and {from} the fact that the
components of $F$ that enter $W$ have always two legs inside $X_4$, we
obtain by a dimensional reduction over $X_4$ an effective Chern-Simons
model as discussed in \cite{034,028}. This Chern-Simons model lives on
the 3-manifold identified by the three vectors $(v,u,w)$, which
becomes the worldvolume of D6-branes if $X_4$ has NUT fixed points
(i.e.\ ADE-type singularities).

To summarize, we derived superpotentials from flux compactifications
if the structure group of the internal manifold is G$_2$, SU(3) or
SU(2) and all three cases yield generically $N$=1 vacua in four
dimensions.  The reduction of the structure group was related to
additional spinors on $X_7$, which in turn implied specific vector
fields specifying the superpotential. For SU(3) it is a single vector
field, which gives a foliation of $X_7$ by a 6-manifold $X_6$ and on
this 6-manifold one can define a holomorphic 3-form that enters the
superpotential.  If the structure group is only SU(2), three vector
fields\footnote{Any 7-dimensional spin manifold has three vector
fields \cite{380} and hence one can always define SU(2) structures.}
define a fibration of a 3-space over a 4-manifold $X_4$.  To get
contact with the expressions discussed in \cite{034,028}, one can
reduce the SU(3) case over $X_6$ to obtain an effective 1-d matrix
model description and for the SU(2) case a reduction over $X_4$ would
gives rise to a 3-d Chern-Simons model.  This is an interesting
observation which requires however further investigations.

{\bf Acknowledgments} 
The work of K.B. is supported by a Heisenberg
grant of the DFG and the work of C.J. by a Graduiertenkolleg grant of
the DFG (The Standard Model of Particle Physics - structure, precision
tests and extensions).



%

\providecommand{\href}[2]{#2}\begingroup\raggedright\endgroup


\end{document}